\documentclass[showpacs,prd,preprintnumbers,nofootinbib]{revtex4}%
\usepackage{amssymb}
\usepackage[T1]{fontenc}
\usepackage{amsmath}
\usepackage{graphicx}
\usepackage{hyperref}
\usepackage{color}
\usepackage{amsfonts}%
\providecommand{\U}[1]{\protect\rule{.1in}{.1in}}

\newcommand{\be}{\begin{equation}}
\newcommand{\ee}{\end{equation}}

\newif\ifinclurecommentaire
\inclurecommentairefalse

\begin{document}
\title{Near-BPS Skyrmions: \\Non-shell configurations and Coulomb effects}
\author{Eric Bonenfant, Louis Harbour and Luc Marleau}
\affiliation{D\'{e}partement de Physique, de G\'{e}nie Physique et d'Optique,
Universit\'{e} Laval, Qu\'{e}bec, Qu\'{e}bec, Canada G1K 7P4}
\date{\today}

\begin{abstract}
The relatively small binding energy in nuclei suggests that they may be well
represented by near-BPS Skyrmions since their mass is roughly proportional to
the baryon number $A.$ For that purpose, we propose a generalization of the
Skyrme model with terms up to order six in derivatives of the pion fields and
treat the nonlinear $\sigma$ and Skyrme terms as small perturbations. For our
special choice of mass term (or potential) $V$, we obtain well-behaved
analytical BPS-type solutions with nonshell configurations for the baryon
density, as opposed to the more complex shell-like configurations found in
most extensions of the Skyrme model . Along with static and (iso)rotational
energies, we add to the mass of the nuclei the often neglected Coulomb energy
and isospin breaking term. Fitting the four model parameters, we find a
remarkable agreement for the binding energy per nucleon $B/A$ with respect to
experimental data. These results support the idea that nuclei could be
near-BPS Skyrmions.

\end{abstract}

\maketitle

%

\ifinclurecommentaire

nomdufichiernotesarticle = pubnotes2012PRD1v3.tex%

\fi

\section{\label{sec:Intro}Introduction}

The idea suggested by Skyrme \cite{Skyrme} that baryon physics could emerge as
solitons from an effective Lagrangian of meson fields remains one of the most
original and successful attempts for the description of the low-energy regime
of the theory of strong interactions (QCD). Although it predates QCD and was
almost eclipsed by it, the proposal gained strong support when it was realized
that, in the large $N_{c}$ limit, QCD is equivalent to an effective theory of
mesons \cite{thooft,Witten}. Perhaps the most important feature of the Skyrme
model in that regard is that the soliton solutions which arise are
characterized by a conserved topological charge, the winding number, which
Skyrme identified as the baryon number. In other words, in this scheme, the
baryons as well as nuclei are simply topological solitons.

In its original formulation, the Skyrme model succeeds in predicting the
properties of the nucleon within a precision of 30\%. This is considered a
rather good agreement for a two-parameter theory \cite{AdkinsWitten}. However,
a number of generalizations of the model have been proposed to improve this
concordance with baryon and nuclear physics. They mostly exploit our ignorance
of the exact form of the low-energy effective Lagrangian of QCD for example,
the structure of the mass term \cite{Marleau1,KPZ,Marleau2}, the contribution
of other vector mesons \cite{Sutcliffe, Adkins2} or simply the addition of
higher-order terms in derivatives of the pion fields \cite{Marleau1}.
Unfortunately, for now, QCD alone only gives hints that such extensions should
appear and the complete determination of the effective Skyrme-like Lagrangian
remains a most serious challenge.

Despite such efforts, one of the recurring problems of Skyrme-like Lagrangians
is that they almost inevitably lead to large binding energy for nuclei already
at the classical level. A solution may be at hand by constructing effective
Lagrangians with soliton solutions that saturate the Bogomol'nyi bound, i.e.
so-called Bogomol'nyi-Prasad-Sommerfield type (BPS) Skyrmions, since their
classical static energy grows linearly with the baryon number $A$ (or atomic
number) much like the nuclear mass. Support for this idea comes from a recent
result from Sutcliffe \cite{Sutcliffe2} who found that BPS-type Skyrmions seem
to emerge for the original Skyrme model when a large number of vector mesons
are added. The additional degrees of freedom cause the mass of the soliton to
decrease down to the saturation of the Bogomol'nyi bound. A different and more
direct approach was proposed by Adam, Sanchez-Guillen, and Wereszczynski (ASW)
\cite{Adam} by means of a prototype model consisting of only two terms: one of
order six in derivatives of the pion fields \cite{Jackson6th} and a second
term, called the potential, which is chosen to be the customary mass term for
pions in the Skyrme model \cite{Adkins}. The model leads to BPS-type compacton
solutions with size and mass growing as $A^{\frac{1}{3}}$ and $A$
respectively, a result in general agreement with experimental observations.
However, the connection between the ASW model and pion physics, or the Skyrme
model, is more obscure due to the absence of the nonlinear $\sigma$ and
so-called Skyrme terms which are of order 2 and 4 in derivatives, respectively.

Following this picture, some of us \cite{Marleau5} have reexamined a more
realistic generalization of the Skyrme model which includes terms up to order
six in derivatives \cite{Jackson6th} in the sector where the nonlinear
$\sigma$ and Skyrme terms are small. In that limit and for an appropriate
choice of mass term, it is possible to find well-behaved analytical solutions
for the static solitons. Since they saturate the Bogomol'nyi bound, their
static energy is directly proportional to $A$ and we recover some of the
results in Ref. \cite{Adam}. In fact, these solutions allow computing
analytically the mass of the nuclei (static and rotational energy) in the
regime where quadratic and quartic terms are small perturbations. Adjusting
the four parameters of the model to fit the resulting binding energies per
nucleon with respect to the experimental data of the most abundant isotopes
leads to an impressive agreement.

These results support the idea of a BPS-type Skyrme model as the dominant
contribution to an effective theory for the properties of nuclear matter.
However, a few issues remain to be addressed before such a model is considered
viable. We shall concentrate on two of them in this work. First, as for most
extensions of the Skyrme model, the BPS-type models in Refs. \cite{Adam} and
\cite{Marleau5} generate shell-like configurations for the energy and baryon
densities as opposed to what experimental data suggests, i.e. almost constant
densities in the nuclei. We show here that it is possible to construct an
effective Lagrangian which leads to nonshell configurations and still
preserves the agreement with nuclear data. The second issue concerns the
inclusion of the Coulomb energy and the isospin symmetry breaking term in the
calculation of nuclear masses. In the context of the Skyrme model, these
contributions have been thoroughly studied for $A=1$
\cite{Durgut:1985mu,Kaelbermann:1986ne,Ebrahim:1987mu,Jain:1989kn,Weigel:1989eb,Rathske:1988qt,Meissner:2009hm}
but are usually neglected, to a first approximation, for higher $A$ since they
are not expected to overcome the binding energies which are usually large and
also because finding the configurations is already numerically challenging so
that only small $A$ solutions are known (e.g. approximate toroidal,
tetrahedral, cubic configurations for $A=1,2,3$ standard Skyrmions,
respectively). However, for our type of near-BPS model, they may have a
significant impact on the predictions given the already good agreement with
data. Moreover, it turns out that the calculation of the Coulomb energy is
simplified by the axial symmetry of the solutions and is calculable for all
$A$.

\section{\label{sec:Skyrme}The near-BPS Skyrme model}

We propose to study the model based on the Lagrangian density
\begin{equation}
\mathcal{L}=\mathcal{L}_{0}+\mathcal{L}_{2}+\mathcal{L}_{4}+\mathcal{L}_{6}
\label{model0to6}%
\end{equation}
with
\begin{align}
\mathcal{L}_{0}  &  =-\mu^{2}V(U)\label{L0}\\
\mathcal{L}_{2}  &  =-\alpha\text{Tr}\left[  L_{\mu}L^{\mu}\right]
\label{L2}\\
\mathcal{L}_{4}  &  =\beta\text{Tr}\left(  \left[  L_{\mu},L_{\nu}\right]
^{2}\right) \label{L4}\\
\mathcal{L}_{6}  &  =-\frac{3}{2}\frac{\lambda^{2}}{16^{2}}\text{Tr}\left(
\left[  L_{\mu},L_{\nu}\right]  \left[  L^{\nu},L^{\lambda}\right]  \left[
L_{\lambda},L_{\mu}\right]  \right)  \label{L6}%
\end{align}
where $L_{\mu}=U^{\dagger}\partial_{\mu}U$ is the left-handed current of the
meson fields represented by the $SU(2)$ matrix $U=\phi_{0}+i\tau_{i}\phi_{i}$
which obey the nonlinear condition $\phi_{0}^{2}+\phi_{i}^{2}=1.$

The constants $\mu,$ $\alpha$, $\beta,$ and $\lambda$ are left as free
parameters of the model although we shall be interested in the regime where
$\alpha$ and $\beta$ are small. The original Skyrme model was built out of
only the nonlinear $\sigma$ term, $\mathcal{L}_{2},$ and the Skyrme term,
$\mathcal{L}_{4}$. One often adds the so-called mass term, $\mathcal{L}_{0},$
to take into account chiral symmetry breaking and generate a pion mass term
for small fluctuations of the chiral field in $V(U)$. Finally, the term of
order six in derivatives of the pion fields $\mathcal{L}_{6}$ is equivalent
to
\[
\mathcal{L}_{J6}=-\frac{\varepsilon_{6}}{4}\mathcal{B}^{\mu}\mathcal{B}_{\mu}%
\]
that was first proposed by Jackson et al. \cite{Jackson6th} to allow for the
possibility of $\omega$-meson interactions. Here, we define the topological
(baryon) current density $\mathcal{B}^{\mu}:$
\begin{equation}
\mathcal{B}^{\mu}=\frac{\epsilon^{\mu\nu\rho\sigma}}{24\pi^{2}}\text{Tr}%
\left(  L_{\nu}L_{\rho}L_{\sigma}\right)  .
\end{equation}
The boundary condition at infinity must be constant to ensure that solutions
for the Skyrme field have finite energy but it also characterizes solutions by
a conserved topological charge,
\begin{equation}
A=\int d^{3}r\mathcal{B}^{0}=-\frac{\epsilon^{ijk}}{24\pi^{2}}\int
d^{3}r\text{Tr}\left(  L_{i}L_{j}L_{k}\right)  .
\end{equation}
The static energy arising from $\mathcal{L}_{6}$ comes from the square of the
baryon density%
\[
E_{6}=\frac{\varepsilon_{6}}{4}\int\left(  \mathcal{B}^{0}\left(
\mathbf{r}\right)  \right)  ^{2}d^{3}r
\]
so in a sense, it is the analog of the Coulomb energy
\begin{equation}
E_{C}=\frac{1}{2}\int\frac{\rho\left(  \mathbf{r}\right)  \rho\left(
\mathbf{r}^{\prime}\right)  }{4\pi\left\vert \mathbf{r}-\mathbf{r}^{\prime
}\right\vert }d^{3}rd^{3}r^{\prime} \label{ECoulomb}%
\end{equation}
except that instead of following the $\left\vert \mathbf{r}-\mathbf{r}%
^{\prime}\right\vert ^{-1}$ law, the interaction is replaced by a $\delta
-$function%
\[
E_{6}=\frac{\varepsilon_{6}}{4}\int\mathcal{B}^{0}\left(  \mathbf{r}\right)
\mathcal{B}^{0}\left(  \mathbf{r}^{\prime}\right)  \delta^{3}\left(
\mathbf{r}-\mathbf{r}^{\prime}\right)  d^{3}rd^{3}r^{\prime}.
\]
In other words, the baryonic charge interacts locally.

Historically, $\mathcal{L}_{0}$ and $\mathcal{L}_{6}$ were introduced to
provide a more general effective Lagrangian than the original Skyrme model and
indeed, the Lagrangian in (\ref{model0to6}) represents the most general
$SU(2)$ model with at most two time derivatives. As an effective theory based
on the $1/N_{c}$ expansion of QCD, there no reason to believe that
higher-order derivatives should be absent. However, since one generally relies
on the standard Hamiltonian interpretation for the quantization procedure,
higher-order time derivatives are usually avoided.

As a result, the model has been studied extensively but remarkably, this was
done only for values of parameters $\mu,$ $\alpha$, $\beta,$ and $\lambda$
close to that of the original Skyrme model. Presumably these choices were made
so that $\mathcal{L}_{2}$ and $\mathcal{L}_{4}$ would continue to have a
significant contribution to the mass of the baryons and thereby preserve the
relative successes of the Skyrme model in predicting nucleon properties and
their link to pion physics ($\alpha$ is proportional to the pion decay
constant $F_{\pi}$). Yet this sector of the theory fails to provide an
accurate description of the binding energy of heavy nuclei.

Noting that this caveat may come from the fact that the solitons of the Skyrme
model do not saturate the Bogomol'nyi bound, ASW proposed a model \cite{Adam}
(equivalent to setting $\alpha=\beta=0)$ whose solutions are BPS-type solitons
and have lower binding energies. A more realistic approach was proposed in
Ref. \cite{Marleau5} to analyze the full Lagrangian (\ref{model0to6}) in the
sector where $\alpha$ and $\beta$ are relatively small treating these two
terms as perturbations. However, in spite of a very good agreement with
experimental nuclear masses, there remains an obstacle to the acceptance of
such model. Nuclear matter is believed to be uniformly distributed inside a
nucleus whereas the solutions of the aforementioned models display shell-like
configuration for the baryon and energy densities. Part of this work is to
demonstrate that it is possible to construct an effective Lagrangian which
leads to nonshell configuration and still preserves and even improves the
agreement with nuclear mass data.

We may write the general static solution as
\begin{equation}
U=e^{i\mathbf{n}\cdot\mathbf{\tau}F}=\cos F+i\mathbf{n}\cdot\mathbf{\tau}\sin
F \label{Hedgehog}%
\end{equation}
where $\mathbf{\hat{n}}$ is the unit vector
\begin{equation}
\mathbf{\hat{n}}=\left(  \sin\Theta\cos\Phi,\sin\Theta\sin\Phi,\cos
\Theta\right)
\end{equation}
Let us consider the model in (\ref{model0to6}) in the limit where $\alpha$ and
$\beta$ are small. For that purpose, we introduce the axial solutions for the
$\alpha=\beta=0$ case$,$
\begin{equation}
F=F(r),\qquad\Theta=\theta,\qquad\Phi=n\phi\label{axialsolution}%
\end{equation}
where $n$ is an integer. The static energy arising from \ref{model0to6}
becomes
\begin{align}
E_{\text{stat}}  &  =4\pi\int r^{2}dr\left(  \mu^{2}V+\frac{9\lambda^{2}}%
{16}n^{2}F^{\prime}{}^{2}\frac{\sin^{4}F}{r^{4}}\right. \nonumber\\
&  +2\alpha\left[  F^{\prime2}+\left(  n^{2}+1\right)  \frac{\sin^{2}F}{r^{2}%
}\right] \nonumber\\
&  +\left.  16\beta\frac{\sin^{2}F}{r^{2}}\left[  \left(  n^{2}+1\right)
F^{\prime2}+n^{2}\frac{\sin^{2}F}{r^{2}}\right]  \right)  \label{Estat}%
\end{align}
Here $F^{\prime}=\partial F/\partial r$ and the topological charge simplifies
to
\begin{equation}
A=-\frac{2n}{\pi}\int F^{\prime}\sin^{2}Fdr=n \label{Bint}%
\end{equation}
Minimizing of the static energy for $\alpha=\beta=0$ leads to the differential
equation for $F:$
\begin{equation}
\frac{9\lambda^{2}n^{2}}{4}\frac{\sin^{2}F}{2r^{2}}\partial_{r}\left(
\frac{\sin^{2}F}{r^{2}}F^{\prime}\right)  -\mu^{2}\frac{\partial V}{\partial
F}=0. \label{minimisation}%
\end{equation}
The change of variable $z=\frac{2\sqrt{2}\mu r^{3}}{9n\lambda}$ allows this
last expression to be written in a simple form
\begin{equation}
\sin^{2}F\partial_{z}\left[  \sin^{2}F\left(  \partial_{z}F\right)  \right]
-\frac{\partial V}{\partial F}=0
\end{equation}
that can be integrated
\begin{equation}
\frac{1}{2}\sin^{4}F\left(  F_{z}\right)  ^{2}=V. \label{equipartition}%
\end{equation}
Regrouping the terms, we get
\begin{equation}
\int dF\frac{\sin^{2}F}{\sqrt{2V}}=\pm\left(  z-z_{0}\right)  \label{gdeu}%
\end{equation}
where $z_{0}$ is an integration constant. Finally, the expression for $F(z)$
can be found analytically provided the integral on the left-hand side is an
invertible function of $F$.

The\ potential (or so-called mass term) $V$ in (\ref{gdeu}) is a key
ingredient in the determination of the solution here. Unfortunately, its exact
form is unknown and indeed, has been the object of several discussions
\cite{Adkins,Marleau1,KPZ}. For simplicity, it is often assumed to be%
\[
V=\frac{1}{4}\text{Tr}\left[  U+U^{\dagger}-2\right]  =1-\cos F.
\]
This form was considered in ASW for $\alpha=\beta=0$ in the context of
BPS-Skyrmions and solving (\ref{gdeu}) for $F$ led to a BPS-compacton
$F(r)=2\arccos\left(  \nu^{1/3}r\right)  $ for $r\in\left[  0,\nu^{-\frac
{1}{3}}\right]  ,$ where $\nu=\frac{\mu}{18n\lambda}$ is a constant depending
on the parameters $\lambda$, $\mu,$ and $n$. Note that $F^{\prime}$ diverges
as $\ r\rightarrow\nu^{-\frac{1}{3}}$ and vanishes at $r=0$. Since this
solution saturates the Bogomol'nyi bound, the static energy is proportional to
the baryon number $A=n$.

A more general choice was introduced in Ref. \cite{Marleau1}:
\begin{equation}
-\mu^{2}V=\sum_{k=1}^{4}C_{k}\text{Tr}\left[  U^{k}+U^{\dagger k}-2\right]
\label{ck}%
\end{equation}
This form allows one to recover the chiral symmetry breaking pion mass term
$-\frac{1}{2}m_{\pi}^{2}\mathbf{\pi}\cdot\mathbf{\pi}$ \ in the limit of small
pion field fluctuations $U=2\exp(i\tau_{a}\pi_{a}/F_{\pi})$ and \ to find a
relation between the pion mass $m_{\pi}$ and the parameter $\mu,$
\begin{equation}
\sum_{k=1}^{\infty}k^{2}C_{k}=-\frac{m_{\pi}^{2}F_{\pi}^{2}}{16}.
\label{mumpi}%
\end{equation}
The case considered in Ref. \cite{Marleau5} is a particular example of such
potential with%
\begin{equation}
-C_{1}=C_{2}=C_{3}=4C_{4}=\frac{\mu^{2}}{128}.
\end{equation}
and $C_{k>4}=0.$ Assuming the axial solution (\ref{axialsolution}), the
potential simplifies to
\begin{equation}
V=\sin^{2}\left(  \frac{F}{2}\right)  \cos^{6}\left(  \frac{F}{2}\right)  .
\label{Vr3}%
\end{equation}
and upon integration (\ref{gdeu}), we get the solution
\begin{equation}
F(r)=\mp2\left\vert \arccos\left(  e^{-\nu r^{3}}\right)  \right\vert
\label{BMsoln}%
\end{equation}
with $\nu=\frac{\mu}{18n\lambda}$. Here, we use the absolute value in order to
eliminate of the sign ambiguity of the arccos function. In order to set the
baryon number to $\left\vert A\right\vert =n$ and the integration constant
$z_{0},$ we fix the boundary conditions $F(0)=0$ and $F(\infty)=\mp\pi$ for
positive and negative baryon number respectively. Note that the exponential
fall off of $F$ at large $r$ helps prevent some quantities such as the moments
of inertia from becoming infinite.

Unfortunately, the BPS-type models in Refs. \cite{Adam} and \cite{Marleau5}
both lead to shell configurations for the baryon and energy densities which
disagrees with experimental results. This is often the case for Skyrme models
and it is clear from expressions (\ref{Estat}) and (\ref{Bint}) that this
behavior can be traced back to the form of the profile $F(r)$ or more
precisely to the derivative $F^{\prime}(r)$ which tends to zero near $r=0$ for
such models.

Let us consider the more appropriate solution of the form
\begin{equation}
F(r)=\mp2\left\vert \arccos\left(  e^{-ar^{2}}\right)  \right\vert
\label{BHMsoln}%
\end{equation}
with $a=\nu^{\frac{2}{3}}$ and similar boundary conditions $F(0)=0$ and
$F(\infty)=\mp\pi$ \ Here, since $F^{\prime}(0)$ $\neq0,$ neither the baryon
density
\begin{equation}
\mathcal{B}^{0}(r)=-\frac{n}{2\pi^{2}}\frac{\sin^{2}F}{r^{2}}F^{\prime}%
=\frac{2an}{\pi^{2}}\frac{e^{-ar^{2}}}{r}\sqrt{1-e^{-2ar^{2}}} \label{B0}%
\end{equation}
nor the static energy density vanishes near $r=0$. We find by inspection of
(\ref{gdeu}), that this solution emerges from a potential similar to
(\ref{Vr3}), namely
\begin{equation}
V=-\frac{8\sin^{2}\left(  \frac{F}{2}\right)  \cos^{6}\left(  \frac{F}%
{2}\right)  }{9\ln\left(  \cos^{2}\left(  \frac{F}{2}\right)  \right)  }.
\label{Vr2}%
\end{equation}
The logarithmic dependence in the denominator of this expression could be
problematic since $F=0$ at $r=0$ and $F=\mp\pi$ \ at $r=\infty$ but the limits
for $V$ are well defined and finite, i.e. $\lim_{r\rightarrow0,\infty}%
V=\frac{8}{9},0$ respectively. It is interesting to note that according to
(\ref{equipartition}), the square root of the potential%
\begin{equation}
\sqrt{V}=\frac{3\lambda\pi}{8\mu}\left(  -\frac{2n}{\pi}\frac{\sin^{2}F}%
{r^{2}}F^{\prime}\right)  =\frac{3\lambda\pi}{8\mu}\mathcal{B}^{0}(r)
\end{equation}
corresponds to the baryon radial density (the term in parenthesis) up to a
multiplicative constant. Thus, in order to obtain a nonshell baryon density,
it suffices to construct a potential $V$ that does not vanish at small $r.$
Such a potential would also imply that $F^{\prime}(0)\neq0.$ \ Our choice of
potential clearly verifies this requirement but this relation also explains
why the earlier BPS-type models could not generate a nonshell configuration,
namely $V\sim(1-\cos\left(  \frac{F}{2}\right)  )$ and $F^{\prime}(0)=0$ in
that limit.

The expression (\ref{Vr2}) only applies to the axial solution
(\ref{axialsolution}) and we need to write a more general form for $V$ in
terms of $U$ if this is to be used in the expression for the Lagrangian. A
simple but not unique approach to construct the potential is to identify
$\cos(F/2)$ to the expression
\[
\frac{1}{4}\left(  2I+U+U^{\dagger}\right)  =\cos^{2}\left(  \frac{F}%
{2}\right)  I
\]
where $I$ is the identity matrix. Then, a convenient expression for $V$ is
given by
\begin{equation}
-\mu^{2}V(U)=\frac{\mu^{2}}{576}\text{Tr}\left[  \frac{\left(  2I-U-U^{\dagger
}\right)  \left(  2I+U+U^{\dagger}\right)  ^{3}}{\ln\left(  \left(
2I+U+U^{\dagger}\right)  /4\right)  }\right] \nonumber
\end{equation}
Comparing this expression to (\ref{ck}) allows retrieving each coefficient%
\begin{align*}
C_{1}  &  =-0.129631\mu^{2},\quad C_{2}=-0.100632\mu^{2},\quad C_{3}%
=-0.045532\mu^{2},\\
C_{4}  &  =-0.0105061\mu^{2},...
\end{align*}
such that
\[
\sum_{k=1}^{\infty}k^{2}C_{k}=-1.12798\mu^{2}=-\frac{m_{\pi}^{2}F_{\pi}^{2}%
}{16}.
\]

Inserting expression (\ref{BHMsoln}) in (\ref{Estat}), we get the static
energy of the soliton in the small $\alpha$ and $\beta$ approximation%
\[
E_{\text{stat}}=E_{0}+E_{2}+E_{4}+E_{6}%
\]

with%
\begin{align}
E_{0}  &  =\frac{4}{3}\sqrt{2}\left(  -3+2\sqrt{3}\right)  n\pi^{3/2}%
\lambda\mu\nonumber\\
E_{2}  &  =\left(  8\left(  \sqrt{2}-1\right)  \left(  n^{2}+1\right)
+6\sqrt{2}\zeta\left(  \frac{5}{2}\right)  \right)  \pi^{3/2}\alpha\nu
^{-1/3}\label{Estatn}\\
E_{4}  &  =64\left(  2\left(  16\sqrt{2}\left(  \sqrt{3}-1\right)  -15\right)
n^{2}+2\right)  \pi^{3/2}\beta\nu^{1/3}\nonumber\\
E_{6}  &  =\frac{4}{3}\sqrt{2}\left(  -3+2\sqrt{3}\right)  n\pi^{3/2}%
\lambda\mu\nonumber
\end{align}
where $\nu=\frac{\mu}{18n\lambda}$ sets the scale of the solution and $\zeta$
is the Riemann $\zeta-$ function . The terms $V$ and $E_{6}$ are proportional
to the baryon number $A=n$ as one expects from solutions that saturate the
Bogomol'nyi bound whereas the small perturbations $E_{2}=A^{1/3}(a_{2}%
+b_{2}A^{2})$ and $E_{4}=A^{-1/3}(a_{4}+b_{4}A^{2})$ have a more complex
dependence. Part of this behavior, the overall factor $A^{\pm1/3},$ is due to
the scaling. The additional factor of $A^{2}$ comes from the axial symmetry of
the solution (\ref{axialsolution}). Note that it is also easy to calculate
analytically the root mean square radius of the baryon density
\begin{equation}
\left\langle r^{2}\right\rangle ^{\frac{1}{2}}=\frac{1}{2}\left(
\frac{18A\lambda}{\mu}\right)  ^{\frac{1}{3}}\sqrt{-1+\ln16} \label{rrms}%
\end{equation}
which is consistent with experimental observation for the charge distribution
of nuclei $\left\langle r^{2}\right\rangle ^{\frac{1}{2}}=r_{0}A^{\frac{1}{3}%
}$.

In order to represent physical nuclei, we have taken into account their
rotational and isorotational degrees of freedom and quantize the solitons. The
standard procedure is to use the semiclassical quantization which is described
in the next section.

\section{\label{sec:Quantization}Quantization}

Skyrmions are not pointlike particles. So we resort to a semiclassical
quantization method which consists in adding an explicit time dependence to
the zero modes of the Skyrmions and applying a time-dependent (iso)rotations
on the Skyrme fields by $SU(2)$ matrix $A(t)$ and $B(t)$
\begin{equation}
\tilde{U}(\mathbf{r},t)=A(t)U(R(B(t))\mathbf{r})A(t)
\end{equation}
where $R(B(t))$ is the associated $SO(3)$ rotation matrix. The approach
assumes that the Skyrmion behave as a rigid rotator. Upon insertion of this
ansatz in the time-dependent part of the full Lagrangian (\ref{model0to6}), we
can write the (iso)rotational Lagrangian as
\begin{equation}
\mathcal{L}_{\text{rot}}=\frac{1}{2}a_{i}U_{ij}a_{j}-a_{i}W_{ij}b_{j}+\frac
{1}{2}b_{i}V_{ij}b_{j},
\end{equation}
where $a_{i}=-i$Tr$A^{\dag}\dot{A}$ and $b_{i}=i$Tr$\dot{B}B^{\dag}$

The moment of inertia tensors $U_{ij}$ is given by%
\begin{align}
U_{ij}  &  =\int d^{3}r\ \mathcal{U}_{ij}=-\int d^{3}r\left[  2\alpha
\text{Tr}\left(  T_{i}T_{j}\right)  \right. \nonumber\\
&  +4\beta\text{Tr}\left(  \left[  L_{p},T_{i}\right]  \left[  L_{p}%
,T_{j}\right]  \right) \nonumber\\
&  +\left.  \frac{9\lambda^{2}}{16^{2}}\text{Tr}\left(  \left[  T_{i}%
,L_{p}\right]  \left[  L_{p},L_{q}\right]  \left[  L_{q},T_{j}\right]
\right)  \right]  \label{MInertia}%
\end{align}
where $T_{i}=iU^{\dagger}\left[  \frac{\tau_{i}}{2},U\right]  $. The
expressions for $W_{ij}$ and $V_{ij}$ are similar except that the
isorotational operator $T_{i}$ is replaced by a rotational analog
$S_{i}=-\epsilon_{ikl}x_{k}L_{l}$ as follows:
\begin{align}
W_{ij}  &  =\int d^{3}r\ \mathcal{W}_{ij}=\int d^{3}r\ \mathcal{U}_{ij}%
(T_{j}\rightarrow S_{j})\label{Wij}\\
V_{ij}  &  =\int d^{3}r\ \mathcal{V}_{ij}=\int d^{3}r\ \mathcal{U}_{ij}%
(T_{j}\rightarrow S_{j},T_{i}\rightarrow S_{i}). \label{Vij}%
\end{align}
Following the calculations in \cite{Marleau5} for axial solution of the form
(\ref{axialsolution}), we find that all off-diagonal elements of the inertia
tensors vanish. Furthermore, one can show that $U_{11}=U_{22}$ and $U_{33}$
can be obtained by setting $n=1$ in the expression for $U_{11}$. Similar
identities hold for $V_{ij}$ and $W_{ij}$ tensors. The axial symmetry of the
solution imposes the constraint $L_{3}+nK_{3}=0$ which is simply the statement
that a spatial rotation by an angle $\theta$ about the axis of symmetry can be
compensated by an isorotation of $-n\theta$ about the $\tau_{3}$ axis. It
follows from expressions (\ref{MInertia})-(\ref{Vij}) that $W_{11}=W_{22}=0$
for $\left\vert n\right\vert \geq2$ and $n^{2}U_{33}=nW_{33}=V_{33}$.

The general form of the rotational Hamiltonian is given by \cite{Houghton2}
\begin{widetext}
\begin{equation}
H_{\text{rot}}=\frac{1}{2}\left[  \frac{\left(  L_{1}+W_{11}\frac{K_{1}%
}{U_{11}}\right)  ^{2}}{V_{11}-\frac{W_{11}^{2}}{U_{11}}}+\frac{\left(
L_{2}+W_{22}\frac{K_{2}}{U_{22}}\right)  ^{2}}{V_{22}-\frac{W_{22}^{2}}%
{U_{22}}}+\frac{\left(  L_{3}+W_{33}\frac{K_{3}}{U_{33}}\right)  ^{2}}%
{V_{33}-\frac{W_{33}^{2}}{U_{33}}}+\frac{K_{1}^{2}}{U_{11}}+\frac{K_{2}^{2}%
}{U_{22}}+\frac{K_{3}^{2}}{U_{33}}\right]  \label{Hrot}%
\end{equation}
\end{widetext}where ($K_{i}$) $L_{i}$ the body-fixed (iso)rotation momentum
canonically conjugate to $(a_{i}$) $b_{i}$. The expression for the rotational
energy of the nucleon $A=1$ simplifies due to the spherical symmetry
\begin{equation}
E_{\text{rot}}^{N}=\frac{3}{8U_{11}}.
\end{equation}
It is also easy to calculate the rotational energies for nuclei with winding
number $\left\vert n\right\vert \geq2$%
\begin{equation}
H_{\text{rot}}=\frac{1}{2}\left[  \frac{\mathbf{L}^{2}}{V_{11}}+\frac
{\mathbf{K}^{2}}{U_{11}}+\xi K_{3}^{2}\right]
\end{equation}
with%
\[
\xi=\frac{1}{U_{33}}-\frac{1}{U_{11}}-\frac{n^{2}}{V_{11}}%
\]
These momenta are related to the usual space-fixed isospin ($\mathbf{I}$) and
spin ($\mathbf{J}$) by the orthogonal transformations
\begin{equation}
I_{i}=-R(A_{1})_{ij}K_{j}, \label{eq:I}%
\end{equation}%
\begin{equation}
J_{i}=-R(A_{2})_{ij}^{\text{T}}L_{j}. \label{eq:J}%
\end{equation}
According to (\ref{eq:I}) and (\ref{eq:J}), we see that the Casimir invariants
satisfy $\mathbf{K}^{2}=\mathbf{I}^{2}$ and $\mathbf{L}^{2}=\mathbf{J}^{2}$ so
the rotational Hamiltonian is given by
\begin{equation}
H_{\text{rot}}=\frac{1}{2}\left[  \frac{\mathbf{J}^{2}}{V_{11}}+\frac
{\mathbf{I}^{2}}{U_{11}}+\xi K_{3}^{2}\right]  . \label{Erot}%
\end{equation}
We are looking for the lowest eigenvalue of $H_{\text{rot}}$ which depends on
the dimension of the spin and isospin representation of the eigenstate
$|i,i_{3},k_{3}\rangle|j,j_{3},l_{3}\rangle$. For $\alpha=\beta=0,$ we can
show that $\xi$ is negative and we shall assume that this remains true for
small values of $\alpha$ and $\beta$. Then, for a given spin $j$ and isospin
$i$, $\kappa$ must take the largest possible eigenvalue $k_{3}.$ Since
$\mathbf{K}^{2}=\mathbf{I}^{2}$ and $\mathbf{L}^{2}=\mathbf{J}^{2},$ the state
with highest weight is characterized by $k_{3}=i$ and $l_{3}=j$ and since
nuclei are build out of $A$ fermions we must have an isospin $j\leq A/2.$ On
the other hand, the axial symmetry of the static solutions implies that
$k_{3}=-l_{3}/n$ where $n=A.$ But for even $A$ nuclei, $k_{3}$ must be an
integer and $\left\vert l_{3}/n\right\vert \leq\left\vert j/n\right\vert
\leq\left\vert A/\left(  2n\right)  \right\vert =1/2$ so
\[
0\leq\left\vert k_{3}\right\vert \leq\left[  \left\vert \frac{A}%
{2n}\right\vert \right]  =0
\]
Similarly for half-integer spin nuclei, $\left\vert k_{3}\right\vert $ must be
a half-integer so the only possible value is
\[
\frac{1}{2}\leq\left\vert k_{3}\right\vert \leq\left\vert \frac{A}%
{2n}\right\vert =\frac{1}{2}%
\]
Summarizing, if we assume for simplicity that the $\alpha$ and $\beta$ terms
only generate small perturbations, the largest possible eigenvalue $k_{3}$ is
\begin{equation}
\kappa=\max(\left\vert k_{3}\right\vert )=\left\{
\begin{tabular}
[c]{l}%
$0\qquad$for $A=$ even\\
$\frac{1}{2}\qquad$for $A=$ odd
\end{tabular}
\ \ \ \ \ \ \ \ \ \ \right.  . \label{kappa}%
\end{equation}

The lowest eigenvalue of the rotational Hamiltonian $H_{\text{rot}}$ for a
nucleus is then given by \cite{Marleau5}
\begin{equation}
E_{\text{rot}}=\frac{1}{2}\left[  \frac{j(j+1)}{V_{11}}+\frac{i(i+1)}{U_{11}%
}+\xi\kappa^{2}\right]  \label{Erotijk}%
\end{equation}
The spin of the most abundant isotopes is fairly well known. The isospins are
not so well known so we resort to the usual assumption that the most abundant
isotopes correspond to states with lowest isorotational energy. Since
$i\geq\left\vert i_{3}\right\vert $, the lowest value that $i$ can take is
simply $\left\vert i_{3}\right\vert $ where $i_{3}=A/2-Z.$ For example, the
deuteron corresponds to $A=n=2,i=0,$ $j=1,$ and $\kappa=0,$ so the rotational
energy reduces to
\begin{equation}
E_{\text{rot}}^{D}=\frac{1}{V_{11}}.
\end{equation}

The explicit calculations of the rotational energy of nuclei then require only
three moments of inertia which can be found analytically:
\begin{align}
U_{11}  &  =\frac{\pi^{3/2}}{18}\left[  24\left(  2\sqrt{2}-1\right)
\alpha\nu^{-1}\right. \nonumber\\
&  +128\left(  2\sqrt{2}\left(  3-4\sqrt{3}\right)  \left(  3n^{2}+1\right)
+3\left(  12n^{2}+7\right)  \right)  \beta\nu^{-1/3}\label{U112}\\
&  +\left.  3\sqrt{2}\left(  8\sqrt{3}-9\right)  \left(  3n^{2}+1\right)
\lambda^{2}\nu^{1/3}\right] \nonumber
\end{align}

\begin{align}
V_{11}  &  =\frac{\pi^{3/2}}{18}\left[  6\left(  2\sqrt{2}-1\right)  \left(
n^{2}+3\right)  \alpha\nu^{-1}\right. \nonumber\\
&  +32\left(  32\sqrt{2}\left(  3-4\sqrt{3}\right)  n^{2}+3\left(
67n^{2}+9\right)  \right)  \beta\nu^{-1/3}\\
&  +\left.  12\sqrt{2}\left(  8\sqrt{3}-9\right)  n^{2}\lambda^{2}\nu
^{1/3}\right] \nonumber
\end{align}
and $U_{33}=U_{11}(n\rightarrow1)$.

So far, both contributions to the mass of the nucleus, $E_{\text{stat}}$ and
$E_{\text{rot}},$ are charge invariant. Since this is a symmetry of the strong
interaction, it is reflected in the construction of the Lagrangian
(\ref{model0to6}) and one expects that the two terms form the dominant portion
of the mass. However, isotope masses differ by a few percent so this symmetry
is broken for physical nuclei. In the next section, we consider two additional
contributions to the mass, the Coulomb energy associated with the charge
distribution inside the Skyrmion and an isospin breaking term\ that may be
attributed to the up and down quark mass difference.

\section{\label{sec:Coulomb}Coulomb energy and isospin breaking}

Even if we thought of a nucleus as a simple collection of individual protons
and neutrons, there would be a repulsive electromagnetic force between protons
and the process would require energy to bring these charges together. The
result is an increase in the mass of the object by an amount corresponding to
the Coulomb energy. Such an effect is of course also present in the Skyrmions
description of nuclei since the static configuration has non-vanishing charge
density. The electromagnetic and isospin breaking contributions to the mass
have been thoroughly studied for $A=1$, mostly in the context of the
computation of the proton-neutron mass difference
\cite{Durgut:1985mu,Kaelbermann:1986ne,Ebrahim:1987mu,Jain:1989kn,Weigel:1989eb,Rathske:1988qt,Meissner:2009hm}%
, but are usually neglected, to a first approximation, for higher $A$ since
they are not expected to overcome the large binding energies predicted by the
model. There are also practical reasons why they are seldom taken into
account. The higher baryon number configurations of the original Skyrme model
are nontrivial (toroidal shape for $A=2$, tetrahedral for $A=3$, etc.) and
finding them exactly either requires heavy numerical calculations (see for
example \cite{Marleau3}) or some kind of clever approximation like rational
maps \cite{RatMap}. Moreover, the computation of the Coulomb energy is more
challenging in general since it involves two integrations over volume. One can
also argue that the Coulomb energy of Skyrmions is somewhat reduced by
shell-like configurations of the charge densities as opposed to what it would
be for a nearly constant spherical density found in electron scattering
experiments. In our case however, we are interested in a more precise
calculation of the nuclei masses and an estimate of the Coulomb energy is
desirable, and even more so in our model which generates nonshell
configurations. It turns out that the analytical form of the chiral angle
$F(r)$ in (\ref{BHMsoln}) and the axial symmetry of the solution simplify the
computation of the Coulomb energy.

Let us first consider the charge density inside Skyrmions. Following Adkins et
al. \cite{AdkinsWitten}, we write the electromagnetic current
\begin{equation}
J_{EM}^{\mu}=\frac{1}{2}\mathcal{B}^{\mu}+J_{V}^{\mu3},
\end{equation}
with $\mathcal{B}^{\mu}$ the baryon density and $J_{V}^{\mu3}$ the vector
current density, so the conserved electric charge is given by
\begin{equation}
Z=\int d^{3}rJ_{EM}^{0}=\int d^{3}r\left(  \frac{1}{2}\mathcal{B}^{0}%
+J_{V}^{03}\right)  =\frac{A}{2}+k_{3} \label{charge}%
\end{equation}
with $k_{3},$ the eigenvalue of third component of isospin in the body-fixed
frame. The vector current is then defined as the sum of the left and right
handed currents
\[
J_{V}^{\mu i}=J_{R}^{\mu i}+J_{L}^{\mu i}%
\]
which are invariant under $SU(2)_{L}\otimes SU(2)_{R}$ transformations of the
form $U\rightarrow LUR^{\dagger}.$ More explicitly, we get%
\begin{equation}
J_{V}^{0i}=\mathcal{U}_{ij}a_{j}-\mathcal{W}_{ij}b_{j} \label{J3V}%
\end{equation}
where $\mathcal{U}_{ij}$ and $\mathcal{W}_{ij}$ are the moment of inertia
densities in (\ref{MInertia})-(\ref{Vij}). In the quantized version, $a_{j}%
\ $and $b_{j}$ are expressed in terms of the conjugate operators $K_{i}$ and
$L_{i}.$ Here we only need the relation
\[
K_{i}=U_{ij}a_{j}-W_{ij}b_{j}%
\]
Since the off-diagonal elements of $U_{ij}$ and $W_{ij}$ vanish when the
solution is axially symmetric and also $n^{2}U_{33}=nW_{33}=V_{33}$, we have
\[
a_{3}=\frac{K_{3}+W_{33}b_{3}}{U_{33}}=\frac{K_{3}}{U_{33}}+nb_{3}%
\]
Inserting $a_{3}$ in (\ref{J3V}), the isovector electric current density
reduces to
\[
J_{V}^{03}=K_{3}\frac{\mathcal{U}_{33}}{U_{33}}%
\]
where $\mathcal{U}_{33}/U_{33}$ may be interpreted here as a normalized moment
of inertia density for the third component of isospin.\ Finally, the electric
charge density is given by
\begin{equation}
\rho(\mathbf{r})\equiv\frac{1}{2}\mathcal{B}^{0}(\mathbf{r})+i_{3}%
\frac{\mathcal{U}_{33}(\mathbf{r})}{U_{33}} \label{rhocharge}%
\end{equation}
where we have replaced $k_{3}$ by $i_{3}$ using the fact that the charge
density of body-fixed and space-fixed frame only differs by a rotation.

The Coulomb energy stored in a charge distribution $\rho(\mathbf{r})$ takes
the usual form (\ref{ECoulomb}). In practice, unless one considers very simple
configurations, it is not possible to find an analytical expression for the
Coulomb energy. Nonetheless, it is often helpful to expand $\rho(\mathbf{r})$
in terms of normalized spherical harmonics to take care of the angular
integrations
\begin{equation}
\rho(\mathbf{r})=\sum_{l,m}\rho_{lm}(r)Y_{l}^{m\ast}(\theta,\phi). \label{Ylm}%
\end{equation}
Following the approach described in \cite{Carlson}, we define the quantities
\begin{equation}
Q_{lm}(r)=\int_{0}^{r}\ dr^{\prime}r^{\prime l+2}\rho_{lm}(r^{\prime})
\label{Qlm}%
\end{equation}
which, at large distance, are equivalent to a multipole moments of the
distribution. Then, each moment contributes to the Coulomb energy by an
amount
\[
U_{lm}=\frac{1}{2\epsilon_{0}}\int_{0}^{\infty}drr^{-2l-2}|Q_{lm}(r)|^{2}\
\]
and the total Coulomb energy associated to the distribution is given by%
\[
E_{C}=\sum_{l=0}^{\infty}\sum_{m=-l}^{l}U_{lm}%
\]

In our case, the angular dependence of the charge density is rather simple.
The first part is a spherically symmetric contribution%
\[
\mathcal{B}^{0}(r)=-\frac{n}{2\pi^{2}}\frac{\sin^{2}F}{r^{2}}F^{\prime}%
\]
whereas the only non-trivial piece comes from the third moment density
$\mathcal{U}_{33}(\mathbf{r})$ and is proportional to $\sin^{2}(\theta)$
\[
\mathcal{U}_{33}=\left(  4\alpha\sin^{2}F+32\beta\sin^{2}F\left(  F^{\prime
2}+\frac{\sin^{2}F}{r^{2}}\right)  +\frac{9\lambda^{2}}{8}F^{\prime2}\sin
^{2}F\left(  \frac{\sin^{2}F}{r^{2}}\right)  \right)  \sin^{2}\theta
=u_{33}(r)\sin^{2}\theta
\]
The summation (\ref{Ylm}) consists of only two terms
\begin{align*}
\rho_{00}(r)  &  =2\sqrt{\pi}\frac{\mathcal{B}^{0}(r)}{2}+\frac{4\sqrt{\pi}%
}{3}\frac{u_{33}(r)}{U_{33}}i_{3}\\
\rho_{20}(r)  &  =-\frac{4}{3}\sqrt{\frac{\pi}{5}}\frac{u_{33}(r)}{U_{33}%
}i_{3}%
\end{align*}
The expressions for moments $Q_{00}(r)$ and $Q_{20}(r)$ are found by
integrating (\ref{Qlm}) analytically. Finally, we obtain the Coulomb energy by
computing numerically the last remaining integral
\begin{equation}
E_{C}=\frac{1}{2\epsilon_{0}}\int_{0}^{\infty}(|Q_{00}|^{2}r^{-4}+|Q_{20}%
|^{2}r^{-8})\ r^{2}dr \label{EC}%
\end{equation}

The Coulomb energy alone cannot explain the isotope mass difference. This is
particularly evident for $A=1$ where the proton mass is known to be smaller
than that of the neutron although the Coulomb energy alone would suggest
otherwise. On the other hand, isospin is not an exact symmetry, a fact that
may be traced back to the up and down quark mass difference. Several attempts
have been made to modelize the isospin symmetry breaking term within the
Skyrme model \cite{Rathske:1988qt,Meissner:2009hm}. Here we shall assume for
simplicity that this results in a contribution proportional to the third
component of isospin%
\begin{equation}
E_{I}=a_{I}i_{3} \label{EI}%
\end{equation}
with the parameter $a_{I}$ fixed by setting the neutron-proton mass difference
to its experimental value. Since both of them have the same static and
rotational energies,%
\[
\Delta M_{n-p}^{\text{expt}}=\left(  E_{C}^{n}-E_{C}^{p}\right)
-a_{I}=1.293\text{ MeV}%
\]
and%
\[
a_{I}=\left(  E_{C}^{p}-E_{C}^{n}\right)  -\Delta M_{n-p}^{\text{expt}}%
\]

Summarizing, the mass of a nucleus reads%
\begin{equation}
E(A,i,j,k_{3},i_{3})=E_{\text{stat}}(A)+E_{\text{rot}}(A,i,j,k_{3}%
)+E_{C}(A,i_{3})+E_{I}(A,i_{3}) \label{Etot}%
\end{equation}
where we have written the explicit dependence of each piece in terms of the
relevant nuclear quantum numbers of the nuclei. The prediction depends on the
parameters of the model $\mu,$ $\alpha,\beta,$ and $\lambda.$

\section{\label{sec:Model}Results and discussion}

The values of the parameters $\mu,$ $\alpha,\beta$ and $\lambda$\ remain to be
fixed. Let us first consider the case where $\alpha=\beta=0.$ This should
provide us with a good estimate for the values of $\mu,\alpha,\beta,$ and
$\lambda$ required in the 4-parameter model (\ref{model0to6}) \ and, after
all, it corresponds to the limit where the minimization of the static energy
leads to the exact analytical solution (\ref{BHMsoln}).

\begin{figure}[ptbh]
\centering\includegraphics[width=0.65\textwidth]{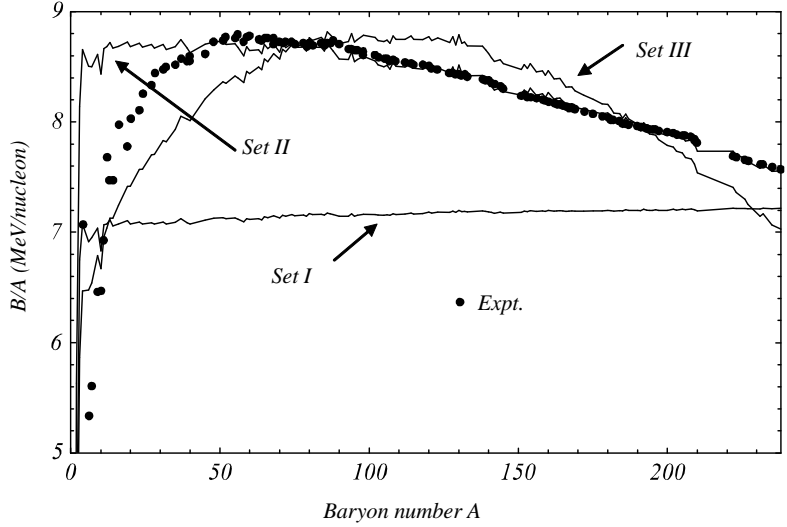}\caption{Binding energy
per nucleon $B/A$ as a function of the baryon number $A$: The experimental
data (black circles) are shown along with predicted values (solid lines) for
parametrization of Set~I ($\alpha=\beta=0$) , Set~II (best fit for nuclear
masses), and Set~III (best fit for $B/A$), respectively.}%
\label{FigBoverA}%
\end{figure}

We need two input parameters to set $\mu$ and $\lambda.$ For simplicity, we
choose the mass of the nucleon and that a nucleus $\ X$ with zero
(iso)rotational energy (i.e. a nucleus with zero spin and isospin) and neglect
for now the Coulomb and isospin breaking energies. The total energy of these
two states is according to (\ref{Estatn}) and (\ref{Erotijk})
\begin{align}
E_{N}  &  =\frac{8}{3}\sqrt{2}\left(  -3+2\sqrt{3}\right)  \pi^{3/2}\lambda
\mu+\frac{\left(  18\right)  ^{4/3}}{32\pi^{3/2}\sqrt{2}\left(  8\sqrt
{3}-9\right)  \left(  \lambda\mu\right)  ^{1/3}\lambda^{4/3}}\\
E_{X}  &  =\frac{8}{3}\sqrt{2}\left(  -3+2\sqrt{3}\right)  n_{X}\pi
^{3/2}\lambda\mu
\end{align}
Solving for $\lambda$ and $\mu$ we get%

\[
\lambda=\frac{3\left(  -3+2\sqrt{3}\right)  ^{1/4}3^{3/4}n_{X}}{4\left(
2E_{X}\right)  ^{1/4}\left(  \pi\left(  8\sqrt{3}-9\right)  \left(  n_{X}%
E_{N}-E_{X}\right)  \right)  ^{3/4}}%
\]%
\begin{equation}
\mu=\frac{\left(  \left(  n_{X}E_{N}-E_{X}\right)  \left(  8\sqrt{3}-9\right)
\right)  ^{3/4}E_{X}^{5/4}}{\left(  3\pi\right)  ^{3/4}\left(  2\left(
2\sqrt{3}-3\right)  \right)  ^{5/4}n_{X}^{2}}%
\end{equation}
As an example, let us examine the case where the nucleus $X$ is Helium-4, the
first doubly magic number nucleus with zero spin and isospin. Setting the mass
of the nucleon as the average mass of the proton and neutron i.e.
$E_{N}=938.919$ MeV and that of Helium-4 nucleus to $E_{He}=3727.38$ MeV, we
get the numerical value $\lambda=0.006\ 413\ 62$ MeV$^{-1}$, $\alpha=\beta=0$
and $\mu=14\ 908.$ MeV$^{2}$ which we shall refer as Set~I. The masses of the
nuclei including static, (iso)rotational, Coulomb, and isospin breaking
contributions are then computed using (\ref{Etot}). Table I shows the relative
deviation of the predicted with regard to experimental values of nuclear
masses of a few isotopes (Set~I). The predictions are accurate to $0.4\%$ or
better even for heavier nuclei. Part of this accuracy is probably due to the
fact that the static energy of a BPS-type solution is proportional to $A$ so
if it dominates, the nuclear masses should follow approximately the same
pattern. However, the predictions remain surprisingly good for a 2-parameter
model$.$ Perhaps more relevant are the predictions of the binding energy per
nucleon ($B/A$). The results are presented in Fig. \ref{FigBoverA} --- Set~I
(solid line) and can be compared to the experimental values (black circles).
We consider here only a subset of the table of nuclei in \cite{Nucltable}
composed of the most abundant 144 isotopes. We observe a sharp rise of the
binding energy per nucleon at small $A$ followed by a slow linear increase for
larger nuclei. The overall accuracy is of the order of $15\%$ which is rather
good considering the fact that the calculation involves the mass difference
between the nucleus and its constituents.

Experimentally the charge radius of the nucleus is known to behave
approximately as
\[
\left\langle r_{\text{em}}^{2}\right\rangle ^{\frac{1}{2}}=r_{0}A^{\frac{1}%
{3}}%
\]
with $r_{0}=1.25$ fm. On the other hand, it is possible to calculate the root
mean square radius for the baryon density [see Eq (\ref{rrms})] \ which leads
to%
\begin{equation}
\left\langle r^{2}\right\rangle ^{\frac{1}{2}}=\left(  2.599\text{ fm}\right)
A^{\frac{1}{3}} \label{r2B}%
\end{equation}
For the charge radius $\left\langle r_{\text{em}}^{2}\right\rangle ^{\frac
{1}{2}}$, the dependence on $A$ is more complex since it involves an
additional isovector contribution (\ref{rhocharge})%
\begin{equation}
\left\langle r_{\text{em}}^{2}\right\rangle =\frac{\int d^{3}rr^{2}%
\rho(\mathbf{r})}{\int d^{3}r\rho(\mathbf{r})}=\frac{A}{2Z}\left\langle
r^{2}\right\rangle +\left\langle r_{V}^{2}\right\rangle \label{r2Z}%
\end{equation}
where $Z=i_{3}+A/2$ is the charge of the nucleus. We get the expression%
\begin{align*}
\left\langle r_{V}^{2}\right\rangle  &  =\frac{i_{3}}{ZU_{33}}\int
drr^{4}u_{33}(r)\\
&  =\frac{i_{3}}{ZU_{33}}\frac{\pi^{3/2}\lambda A^{1/3}}{24\mu^{5/3}}\left[
648\times2^{1/6}3^{1/3}\left(  \sqrt{2}-8\right)  A^{4/3}\alpha\lambda
^{2/3}\right. \\
&  +128\left(  4\sqrt{2}\left(  -9+16\sqrt{3}\right)  -279\right)
A^{2/3}\beta\mu^{2/3}\\
&  +\left.  13\times2^{5/6}3^{1/6}\left(  9\sqrt{3}-32\right)  \lambda
^{4/3}\mu^{4/3}\right]
\end{align*}
where $U_{33}$ also depends on $A$ and is obtained by substituting
$\ U_{33}=U_{11}(n\rightarrow1)$ in (\ref{U112}). Our computation verifies
that the charge radius obeys roughly the proportionality relation
\[
\left\langle r_{\text{em}}^{2}\right\rangle ^{\frac{1}{2}}\sim\left(
2.637\text{ fm}\right)  A^{\frac{1}{3}}%
\]
but overestimates the experimental value of $r_{0}$ by approximately a factor
of 2.

Let us now release the constraints on $\alpha$ and $\beta$ and allow for small
perturbations from the nonlinear $\sigma$ and Skyrme term. In order to
estimate the magnitude of the parameters $\alpha$ and $\beta$ in a real
physical case, we perform two fits: Set~II optimizes the four parameters $\mu
$, $\alpha,$ $\beta$ and $\lambda$ to better reproduce the masses of the
nuclei while Set~III tries to reach the best agreement with respect to the
binding energy per nucleon, $B/A$. Both fits are performed with data from the
same subset of the most abundant 144 isotopes as before. A summary of the
results is presented in Table I while Fig. \ref{FigBoverA} displays the
general behavior of $B/A$ as a function of the baryon number for Sets I, II,
III, and experimental values.%

\[%
\begin{tabular}
[c]{|c|c|c|c|c|}\hline\hline
\multicolumn{5}{|c|}{Table I: Prediction versus experimental nuclear
masses}\\\hline\hline
\  & $\quad$Set~I$\quad$ & $\quad$Set~II$\quad$ & $\quad$Set~III$\quad$ &
Experiment\\\hline\hline
$\mu$ $(10^{4}$ MeV$^{2})$ & $1.490\ 80$ & $1.505\ 71$ & $1.729\ 55$ & \ \\
$\alpha$ $(10^{-3}$ MeV$^{2})$ & $0$ & $5.881\ 18$ & $22.0821$ & \ \\
$\beta$ $(10^{-6}$ MeV$^{0})$ & $0$ & $-1.84877$ & $-5.80989$ & \ \\
$\lambda$ $(10^{-3}$ MeV$^{-1})$ & $6.413\ 62$ & $6.339\ 73$ & $5.536\ 91$ &
\ \\
$F_{\pi}$ $($ MeV$)$ & $0$ & $0.307$ & $0.594$ & $186$\\
$m_{\pi}$ $($MeV$)$ & --- & $208\ 530$ & $82\ 300$ & $138$\\
$e$ $(10^{4})$ & --- & $-185\ 000$ & $-5380$ & \ \\
$r_{0}$ (fm) & $2.637$ & $2.617$ & $2.385$ & $1.23$\\\hline\hline
Nucleus X\  & \multicolumn{3}{|c|}{$\frac{E_{X}-E_{\exp}}{E_{\exp}}$} &
$\ E_{\exp}\text{(MeV)}\ $\\\hline\hline
Nucleon & Input & $-0.0008$ & $0.0020$ & $938.919$\\
$^{2}$H & $-0.0032$ & $-0.0048$ & $-0.0020$ & $1875.61$\\
$^{3}$H & $-0.0042$ & $-0.0057$ & $-0.0030$ & $2808.92$\\
$^{4}$He & Input & $-0.0017$ & $-0.0009$ & $3727.38$\\
$^{6}$Li & $-0.0017$ & $-0.0034$ & $-0.0010$ & $5601.52$\\
$^{7}$Li & $-0.0014$ & $-0.0031$ & $-0.0008$ & $6533.83$\\
$^{9}$Be & $-0.0006$ & $-0.0023$ & $-0.0001$ & $8392.75$\\
$^{10}$B & $-0.0004$ & $-0.0021$ & $-0.00001$ & $9324.44$\\
$^{16}$O & $0.0010$ & $-0.0008$ & $0.0009$ & $14\ 895.1$\\
$^{20}$Ne & $0.0010$ & $-0.0007$ & $0.0008$ & $18\ 617.7$\\
$^{40}$Ca & $0.0016$ & $0.0001$ & $0.0006$ & $37\ 214.7$\\
$^{56}$Fe & $0.0018$ & $0.0001$ & $0.0004$ & $52\ 089.8$\\
$^{238}$U & $0.0004$ & $0.00001$ & $0.0006$ & $221\ 696$\\\hline
\end{tabular}
\ \ \ \ \ \ \ \ \ \ \ \ \ \ \ \ \ \ \ \
\]
We find that the two new sets of parameters are very close to Set
I. The nonlinear $\sigma$ and Skyrme parameters $\alpha$ and $\beta$ are small
in magnitude but in order to make a relevant comparison, it is best to look at
the relative importance of the contributions in (\ref{model0to6}) and how they
scale with respect to the parameters of the model, namely%
\[%
\begin{tabular}
[c]{rccccccc}
& $\lambda\mu$ & $:$ & $\alpha\nu^{-1/3}$ & $:$ & $\beta\nu^{1/3}$ & $:$ &
$\lambda\mu$\\
$\text{Set~I\qquad}$ & $95.61$ & $:$ & $0$ & $:$ & $0$ & $:$ & $95.61$\\
$\text{Set~II\qquad}$ & $95.46$ & $:$ & $4.408\times10^{-5}$ & $:$ &
$-2.255\times10^{-4}$ & $:$ & $95.46$\\
$\text{Set~III\qquad}$ & $95.74$ & $:$ & $1.418\times10^{-4}$ & $:$ &
$-7.904\times10^{-4}$ & $:$ & $95.74$%
\end{tabular}
\ \ \ \ \ \
\]
for $\mathcal{L}_{0},\mathcal{L}_{2},\mathcal{L}_{4},$ and $\mathcal{L}_{6}$
respectively. Clearly, the nonlinear $\sigma$ and Skyrme terms are extremely
small compared to that of $\mathcal{L}_{0}\mathcal{\ }$and $\mathcal{L}_{6},$
i.e. by at least 6 orders of magnitude. This provides support to the
assumption that (\ref{BHMsoln}) is a good approximation to the exact solution.
The overall factor $\lambda\mu$ remains approximately the same for all the
sets but $B/A$ turns out to be somewhat sensitive to these small variations
because it involves a mass difference. Even more sensitive to small change in
parameters is the charge radius $r_{0}$ with $10\%$ decrease between Set~II
and Set~III (Table I) which suggests that the predicted value of $r_{0}$
should be taken as an estimate rather than a firm prediction.

Comparing Set~II and Set~III to the original Skyrme model with a pion mass
term, we may identify%
\[
F_{\pi}=4\sqrt{\alpha},\qquad e^{2}=\frac{1}{32\beta}%
\]
and using (\ref{mumpi}) we find%
\[
m_{\pi}=1.0621\frac{\mu}{\sqrt{\alpha}}.
\]
These quantities, $F_{\pi},e^{2}$ and $m_{\pi}$ take values (see Table I)
which are orders of magnitude away for those obtained for the Skyrme model but
this is expected since we have assumed from the start that $\alpha$ and
$\beta$ are relatively small.

We find also that the Skyrme term has the wrong sign so it would destabilize
the soliton against shrinking if it was not for the contribution of order six
in derivatives which ensures stability against scale transformations. Indeed
the term of order six was even introduced at one point to resolve some
problems with this sign \cite{Jackson6th}. In principle however, a negative
coefficient for the Skyrme term could become problematic since the energy may
no longer be bounded from below. One can argue that for our set of parameters,
the relative weight of the $\mathcal{L}_{4}$ piece with respect to that of
$\mathcal{L}_{0}$ or $\mathcal{L}_{6}$ is so small, i.e. approximately
$\beta\nu^{1/3}/\lambda\mu\sim-10^{-6},$ and is at least partially canceled by
that of the nonlinear $\sigma$ term $\mathcal{L}_{2}$ so that the energy would
remain bounded from below. To substantiate this point on the relative
contribution of each term, it is useful at this point to invoke some relevant
links noticed by Manton \cite{Manton87} between an effective $SU(2)$ scalar
Lagrangian and the strain tensor in the theory of elasticity. As in nonlinear
elasticity theory, the energy density of a Skyrme field depends on the local
stretching associated with the map $U:R^{3}\mapsto S^{3}.$ This is related to
the strain tensor at a point in $R^{3}$ which is defined as
\begin{align*}
M_{ij}  &  =\partial_{i}\Phi\partial_{j}\Phi\ \quad\text{where \quad}%
\Phi=(\sigma,\pi^{z},\pi^{x},\pi^{y})\\
&  =-\frac{1}{4}Tr[\{L_{i},L_{j}\}]
\end{align*}
where $i,j$ refers to the Cartesian space coordinates. $M_{ij}$ is a
$3\times3$ symmetric matrix with three positive eigenvalues $\lambda_{1}%
^{2},\lambda_{2}^{2},$ and $\lambda_{3}^{2}$. Three fundamental invariants
emerges from $M_{ij}$ in this simple geometrical interpretation due to Manton.
They correspond to the Lagrangians $\mathcal{L}_{2},\mathcal{L}_{4},$ and
$\mathcal{L}_{6}$ and lead to the following energy densities, respectively:%

\begin{align}
\mathcal{E}_{1}  &  =\alpha(\lambda_{1}^{2}+\lambda_{2}^{2}+\lambda_{3}%
^{2})\nonumber\\
\mathcal{E}_{2}  &  =-\left\vert \beta\right\vert (\lambda_{1}^{2}\lambda
_{2}^{2}+\lambda_{2}^{2}\lambda_{3}^{2}+\lambda_{1}^{2}\lambda_{3}%
^{2})\label{XYZ}\\
\mathcal{E}_{3}  &  =3\gamma\lambda_{1}^{2}\lambda_{2}^{2}\lambda_{3}%
^{2}\nonumber
\end{align}
where we wrote for simplicity $\gamma=\frac{3}{2}\frac{\lambda^{2}}{16^{2}},$
and \ to the baryon density
\begin{equation}
\mathcal{B}^{0}=\frac{1}{2\pi^{2}}\sqrt{\lambda_{1}^{2}\lambda_{2}^{2}%
\lambda_{3}^{2}} \label{BXYZ}%
\end{equation}
Assuming without loss of generality that $\lambda_{1}^{2}\geq\lambda_{2}%
^{2}\geq\lambda_{3}^{2},$ we find%
\begin{equation}
\mathcal{E}_{1}\geq3\alpha\lambda_{3}^{2},\quad\mathcal{E}_{2}\geq-3\left\vert
\beta\right\vert \lambda_{1}^{2}\lambda_{2}^{2},\quad\mathcal{E}_{3}%
\geq3\gamma\lambda_{1}^{2}\lambda_{2}^{2}\lambda_{3}^{2}\nonumber
\end{equation}
for a total energy density%
\[
\mathcal{E}\geq\mu^{2}V+\frac{\alpha\left\vert \beta\right\vert }{\gamma
}+3\left(  \lambda_{3}^{2}-\frac{\left\vert \beta\right\vert }{\gamma}\right)
\left(  \alpha+\gamma\lambda_{1}^{2}\lambda_{2}^{2}\right)  \geq\mu
^{2}V-3\left\vert \beta\right\vert \left(  \lambda_{1}^{2}\lambda_{2}%
^{2}\right)
\]
So, negative energy density contributions would come from regions where
$\lambda_{3}^{2}<\frac{\left\vert \beta\right\vert }{\gamma}\sim-10^{-6}$, in
other words, where the baryon density $\mathcal{B}^{0}$ in (\ref{BXYZ}) is
very small. Even for $\lambda_{3}^{2}=0$ and $\alpha$ negligibly small, the
energy density should be dominated by the potential term $\mu^{2}V$. If we
consider the integrated energy density subject to the condition that the total
baryon number $A$ is a positive integer, then we expect the energy to be
bounded from below for our set of parameters.

Clearly for our axial solution, the $\mathcal{L}_{2}$ and $\mathcal{L}_{4}$
pieces of the Lagrangian do not play the same significant role in the
stabilization of the soliton as they do in the case of the Skyrme model. The
properties of the soliton are almost completely determined by the values of
$\mu$ and $\lambda$ so $F_{\pi}$ and $m_{\pi}$ may not be so closely related
to the nucleon mass scale as for the original Skyrme model. Perhaps the
explanation for such a departure is that the parameters of the model are
merely bare parameters and they could differ significantly from their
renormalized physical values. In other words, we may have to consider two
quite different sets of parameters: a first one, relevant to the perturbative
regime for pion physics where $F_{\pi}$ and $m_{\pi}$ are close to their
experimental value and a second one, that applies to the nonperturbative
regime in the case of soliton. Unfortunately, one of the successes of the
original Skyrme model is that it established a link between pion physics with
realistic values for $F_{\pi}$ and $m_{\pi}$ and baryon masses. Such a link
here is more obscure.

On the other hand, the model in (\ref{model0to6}) (in the regime where
$\alpha$ and $\beta$ are small) improves the prediction with regard to the
properties of the nuclei of nuclear masses. Let us look more closely at the
results presented in Fig. \ref{FigBoverA}. These are in the form of the ratio
of the binding energy per nucleon ($B/A$) as a function of the baryon (or
atomic) number $A$. The experimental data (black circles) are shown along with
predicted value (solid lines) for parametrization of Set~I , Set~II and Set
III. Set~I is the least accurate when it comes to reproducing the experimental
data, especially in the heavy nuclei sector. Yet, the agreement remains within
a $0.4\%$ of the experimental masses which is much better than with the
original Skyrme model. Moreover, since the ratio $B/A$ depends on the
difference between the mass of a nucleus and that of its constituents, it is
sensitive to small variation of the nuclear masses so the results for $B/A$
may be considered as rather good. The second fit (Set~II), which is optimized
for nuclear masses, overestimates the binding energies of the lightest nuclei
while it reproduces almost exactly the remaining experimental values (
$A\gtrsim40)$. Finally, the least square fit based on $B/A$ (Set~III) is the
best fit overall but in order to provide a better representation for light
nuclei, it abdicates some of the accuracy found in Set~II for $A\gtrsim40$.

This apparent dichotomy between light and heavy nuclei may be partly
attributed to the (iso)rotational contribution to the mass. The size of nuclei
grows as $A^{\frac{1}{3}}$ and their moments of inertia increase accordingly.
Also, the spin of the most abundant isotopes remains small while isospin can
have relatively large values due to the growing disequilibrium between the
number of proton and the number of neutron in heavy nuclei. Our numerical
calculations reveal that the total effect leads to a (iso)rotational energy
$E_{\text{rot}}<$ 1 MeV for $A>10$ for all sets of parameters considered and
its contribution to $B/A$ decreases rapidly as $A$ increases. On the contrary
for $A<10$ the rotational energy is responsible for a larger part of the
binding energy which means that $B/A$ \ should be sensitive to the way the
rotational energy is computed. So clearly, the shape of the baryon density
will have some bearing on the predictions for the small $A$ sector.
\begin{figure}[ptbh]
\centering\includegraphics[width=0.65\textwidth]{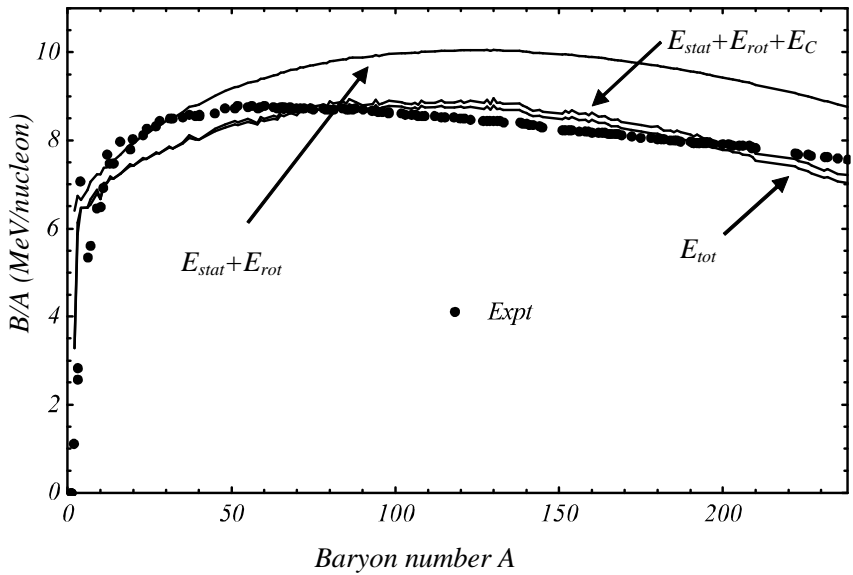}\caption{Contributions
of the Coulomb and isospin effect to the binding energy per nucleon $B/A$ as a
function of the baryon number $A$ for Set~III: The experimental data (black
circles) are shown along with contributions due to $E_{\text{stat}%
}+E_{\text{rot}},$ $E_{\text{stat}}+E_{\text{rot}}+E_{C},$ and $E_{\text{tot}%
}$ =$E_{\text{stat}}+E_{\text{rot}}+E_{C}$ $+E_{I}$ (solid lines).}%
\label{FigCoulomb}%
\end{figure}

Since part of this work is to propose a model with nonshell configuration, it
is relevant to compare our result with a similar analysis \cite{Marleau5}
which involves a typical shell-like configuration. For this purpose we
repeated our calculation omitting the Coulomb and isospin breaking term. It
turns out that both models are equally successful at reproducing data.
Minimizing the square root of the mean squared deviation of $B/A$ from its
experimental value gives almost identical results for both models,
$\sigma=0.50$ MeV per nucleon, despite generating completely different baryon
and energy configurations. In fact, the absence of a variation in $\sigma$
signals somehow the equal inability for both models to provide an accurate
description of light and heavy nuclei sectors at the same time. One could
improve the agreement by fitting separately the parameters $\mu,\alpha,\beta,$
and $\lambda$ in the two sectors $A>40$ and $A<40.$ But this would means
introducing an arbitrary baryon number dependence on the parameter which could
only be justified by introducing some kind of dynamical effect on $\mu
,\alpha,\beta,$ and $\lambda.$

The second motivation for this work regards this addition of the Coulomb and
isospin breaking effect into the nuclear masses. These are often neglected in
the context of the Skyrme model although they must inevitably be taken into
account for a complete description of the nucleus. The Coulomb energy and
isospin breaking term represent small corrections to the nuclear mass (of the
order of $0.1\%$ and $0.01\%,$ respectively) however our results show that the
Coulomb effect is much more significant in the calculation of the binding
energy. The change in $B/A$ is depicted in Fig. \ref{FigCoulomb} for Set~III
in the separation between the red and blue lines (with and without Coulomb
term, respectively). The effect increases almost linearly with the baryon
number up to approximately $2$ MeV per nucleon for the heaviest nuclei. It
represents roughly half the Coulomb effect estimated in the Liquid Drop model
$\left(  B/A\right)  _{\text{Coulomb}}=-a_{C}Z(Z-1)A^{-4/3}$ where the value
of $a_{C}=0.691$ MeV/nucleon. On the other hand, the isospin breaking
contribution due to $E_{I}$ remains very small. Despite the magnitude of these
corrections, it turns out that the optimization of the model parameters only
yield, a slight improvement of the overall agreement with $\sigma=0.48$ MeV
per nucleon.

To summarize, we have proposed a 4-terms model as a generalization of the
Skyrme model. By choosing an appropriate form for the potential $V$, we
allowed for near-BPS solitons with nonshell configurations for the baryon
density in order to achieve a more realistic description of nuclei as opposed
to the more complex configurations found in most extensions of the Skyrme
model (e.g. $A=2$ toroidal , $A=3$ tetrahedral, $A=4$ cubic,...). Moreover, we
introduced additional contributions to the mass of the nuclei coming from the
Coulomb energy and an isospin breaking term. Fitting the model parameters, we
find a remarkable agreement for the binding energy per nucleon $B/A$ with
respect to experimental data. These results suggest that nuclei could be
considered as near-BPS Skyrmions. On the other hand, there remain some
caveats. First, the Skyrme model provides a simultaneous description for
perturbative pion interactions and nonperturbative baryons physics with single
realistic values for $F_{\pi}$ and $m_{\pi}$ and baryon masses. The connection
between the two sectors here seems to be much more intricate. Also, a much
better agreement could be reached if one could construct a solution that would
describe equally well the light and heavy nuclei. Finally, one would like
ultimately to reproduce the observed structure of the nucleus, i.e. a roughly
constant baryon density becoming diffuse at the nuclear surface which is
characterized by a skin thickness parameter. A more appropriate choice of
potential may be instrumental in achieving some of these goals.

This work was supported by the National Science and Engineering Research
Council of Canada.


\begin{thebibliography}{99}                                                                                               %


\bibitem {Skyrme}T.H.R. Skyrme, Proc. Roy. Soc. Lond. A, 260:127-138, 1961;
T.H.R. Skyrme, Proc. Roy. Soc. Lond. A, 247:260--278, 1961; T.H.R. Skyrme,
Nucl. Phys. 31:556-569, 1962; T.H.R. Skyrme, Proc. Roy. Soc. Lond. A,
247:260--278, 1958.

\bibitem {thooft}G. t Hooft, Nuclear Physics B, 72:461-473, 1974.

\bibitem {Witten}E. Witten, Nuclear Physics B, 160:57-115, 1979.

\bibitem {AdkinsWitten}G. S. Adkins,C. R. Nappi and E. Witten, Nucl.Phys.B228:552,1983.

\bibitem {Marleau1}L. Marleau, Phys. Rev. D, 43:885-890, 1991.

\bibitem {Marleau2}E. Bonenfant and L. Marleau, Phys. Rev. D, 80:114018, 2009.

\bibitem {KPZ}V. B. Kopeliovich, B. Piette and W. J. Zakrzewski, Phys. Rev. D,
73:014006, 2006.

\bibitem {Sutcliffe}P. Sutcliffe, Phys. Rev. D, 79:085014, 2009.

\bibitem {Adkins2}G. S. Adkins and C. R. Nappi, Phys. Lett. B, 137:251-256, 1984.

\bibitem {Sutcliffe2}P. Sutcliffe, JHEP 1008:019, 2010.

\bibitem {Adam}C. Adam, J. Sanchez-Guillen and A. Wereszczynski, Phys.Lett.B691:105-110,2010.

\bibitem {Jackson6th}A. Jackson, A. D. Jackson, A. S. Goldhaber, G. S. Brown,
and L. C. Castillo, Phys. Lett. 154B, 101,1985

\bibitem {Adkins}G. S. Adkins and C. R. Nappi, Nucl. Phys. B, 233:109-115, 1984.

\bibitem {Durgut:1985mu}M.~Durgut and N.~K.~Pak,
Phys.\ Lett.\ B \textbf{159}, 357 (1985) [Erratum-ibid.\ \textbf{162B}, 405
(1985)].


\bibitem {Kaelbermann:1986ne}G.~Kalbermann,
Phys.\ Rev.\ C \textbf{34}, 2341 (1986).


\bibitem {Ebrahim:1987mu}A.~Ebrahim and M.~Savci,
Phys.\ Lett.\ B \textbf{189}, 343 (1987).


\bibitem {Jain:1989kn}P.~Jain, R.~Johnson, N.~W.~Park, J.~Schechter and
H.~Weigel,
Phys.\ Rev.\ D \textbf{40}, 855 (1989).


\bibitem {Weigel:1989eb}H.~Weigel, J.~Schechter and N.~W.~Park,
In *Syracuse 1989, Proceedings, 11th Annual Montreal-Rochester-
Syracuse-Toronto Meeting* 110-130. (see Conference Index)

\bibitem {Rathske:1988qt}E.~Rathske,
Z.\ Phys.\ A \textbf{331}, 499 (1988).


\bibitem {Meissner:2009hm}U.~-G.~Meissner, A.~M.~Rakhimov, A.~Wirzba and
U.~T.~Yakhshiev,
EPJ Web Conf.\ \textbf{3}, 06008 (2010) [arXiv:0912.5170 [nucl-th]].


\bibitem {Marleau5}E. Bonenfant and L. Marleau, Phys.Rev. D, 82:054023, 2010.

\bibitem {Houghton2}C. Houghton and S. Magee, Physics Letters B, 632:593-596, 2006.

\bibitem {Marleau3}J.-P. Longpr\'{e} and L. Marleau, Phys. Rev. D, 71:095006, 2005.

\bibitem {RatMap}C.J. Houghton, N.S. Manton and P.M. Sutcliffe, Nucl. Phys.
B510:507, 1998.

\bibitem {Carlson}B.C. Carlson and G.L. Morley, Amer J. Phys. 31:209, 1963.

\bibitem {Nucltable}See for example K. S. Krane, Introductory Nuclear Physics,
John Wiley and sons, p. 67, 1987 or more recent G.Audi, A.H.Wapstra and
C.Thibault, Nuclear Physics A729: 337-676 (2003).

\bibitem {Manton87}N.S. Manton, Commun. Math. Phys. 111, 469 (1987).
\end{thebibliography}
\end{document}